\documentstyle{mn}
\input{epsf}
\def\kms{{\,\rm km\,s^{-1}}}

\begin{document}

\title[]{Disc heating in NGC 2985}

\author[J. Gerssen, K. Kuijken and M. R. Merrifield]{Joris Gerssen$^1$,
Konrad Kuijken$^1$ and Michael R. Merrifield$^2$\\
$^1$Kapteyn Institute, Groningen 9700 AV, The Netherlands\\
$^2$School of Physics \& Astronomy, University of Nottingham \\
}

\maketitle

\begin{abstract}

Various processes have been proposed to explain how galaxy discs acquire
their thickness.  A simple diagnostic for ascertaining this ``heating''
mechanism is provided by the ratio of the vertical to radial velocity
dispersion components.  In a previous paper we have developed a
technique for measuring this ratio, and demonstrated its viability on
the Sb system NGC~488.  Here we present follow-up observations of the
morphologically similar Sab galaxy NGC~2985, still only the second
galaxy for which this ratio has been determined outside of the solar
neighbourhood.  The result is consistent with simple disc heating models
which predict ratios of $\sigma_z / \sigma_R$ less than one. 

\end{abstract}


\section{Introduction}

The three dimensional distribution of velocities within a galactic disc
contains a wealth of information about disc structure.  One aspect that
is immediately tractable with this information is the dynamical history
of a disc.  It has long been known that the random velocity of disc
stars increases over their lifetime, a process dubbed disc heating.  The
two main contributors to this heating process -- {\it i.e.} the increase
of velocity dispersions over time -- are molecular clouds, which scatter
and heat stars more or less isotropically, and spiral irregularities,
which primarily heat stars in the plane of the disc.  Jenkins \& Binney
(1990, see also Jenkins 1992) have numerically studied the combined
effect of these two processes.  By varying the relative importance of
the two mechanisms they showed that the ratio $\sigma_z / \sigma_R$
decreases when the contribution of spiral arm structure increases. 
Although they made their predictions to explain the solar neighbourhood
data, their results can also be applied to other galaxies provided
$\sigma_R$ and $\sigma_z$ can be measured. 

Only in the immediate solar neighbourhood can the full
distribution of stellar positions and velocities be obtained directly. 
The stellar velocities can be fitted by a trivariate Gaussian
\begin{equation}
f({\bf v}) \propto \exp \left[ - \left( \frac{v_R^2}{2 \sigma_R^2} + 
	\frac{ (v_\phi - \overline{v}_\phi)^2}{2 \sigma_\phi^2} 
	+ \frac{v_z^2}{2 \sigma_z^2} \right) \right] ,
\end{equation}
a function originally proposed by Schwarzschild (1907).  Such a
distribution is known as a velocity ellipsoid since the density of stars
is constant on ellipsoids with semi-axes lengths given by the velocity
dispersion components.  Solar neighbourhood observations show that
$\sigma_R > \sigma_\phi > \sigma_z$ which is consistent with the
predictions of the gradual-heating mechanisms described above.  
Other, more erratic heating occurs for instance during a minor merger
event such as the accretion of a small satellite (Sellwood, Nelson \&
Tremaine, 1998).  Depending on the geometry, this irregular heating
process can lead to a substantial thickening of the disc and to a
velocity ellipsoid that is significantly different than in the solar
neighbourhood. 

In a previous paper (Gerssen, Kuijken \& Merrifield 1997, hereafter
Paper~1) we have shown that much can be inferred about the shapes of
the velocity ellipsoids in external galaxies from spectra obtained
along different position angles.  Most studies of stellar velocity
dispersions have however concentrated on systems that are either close
to edge-on or face-on and therefore only provide information about a
single component of the velocity dispersion.  But in an
intermediate-inclination galaxy spectra obtained along radius vectors
with different position angles will show different projections of the
velocity ellipsoid.  In Paper~1 we showed that we can derive the ratio
of the vertical to radial velocity dispersion in NGC~488 from longslit
spectra obtained along the major axis, where the line-of-sight
velocities are a combination of the azimuthal and the vertical
components, and along the minor axis, which gives a combination of the
radial and vertical components.  Along an arbitrarily positioned
spectrum the velocity dispersion in a thin axisymmetric disc can be
written as

\begin{equation}
\sigma_{los}^2=\left[ \sigma_R^2 \sin^2
\phi +\sigma_{\phi}^2 \cos^2 \phi \right] \sin^2 i +\sigma_z^2 \cos^2 i . 
\end{equation}
There are in fact only two independent quantities that can be
extracted from the variation of $\sigma_{los}$ with position angle
(see Paper~1).  Additional spectra, obtained along a third position
angle, will therefore only supply redundant information -- if the disc
is perfectly axisymmetric.  A further constraint is necessary if we
want to retrieve all three components of the velocity dispersion.

Most of the disc stars' orbits are nearly circular and can be adequately
described using the epicycle approximation.  This description leads to a
relation between the radial and azimuthal components of the velocity
dispersion ({\it e.g.} Binney \& Tremaine 1987) which yields the third
constraint, provided that the observed velocity profiles are not
significantly skewed:
\begin{equation}
\frac{\sigma^2_{\phi}}{\sigma^2_{R}} = 
{1 \over 2}\left(1+ {\partial \ln V_c \over \partial \ln R} \right) .
\end{equation}
As shown by Kuijken \& Tremaine (1991), for higher accuracy $V_c$ here
should be taken as the stellar rotation speed within the epicycle
approximation.  Within this approximation, then, to leading order the
shape of the velocity ellipsoid in the plane of the disc is a property
of the galactic potential, independent of any disc heating mechanism. 
Thus, this relation together with longslit spectra obtained along at
least two different position angles (preferably the major and minor axes
for maximum leverage) is sufficient to retrieve all three components of
the stellar velocity ellipsoid.  Cross terms -- {\it i.e.} tilting of
the velocity ellipsoid -- will on average be zero along each line of
sight through a galactic disc. 

\section{Modelling}

The fact that disc stars possess random motions implies that discs are
not completely rotationally supported -- a situation that would be
violently unstable anyway -- and the stellar rotation speed will
therefore be lower than the circular velocity.  The difference between
the two is called the asymmetric drift.  An expression for the
asymmetric drift can be derived by evaluating the velocity moments of
the distribution function $f$, see also paper~1.  We make the assumption
that $\sigma_{Rz}=0$, {\it i.e.} there is no coupling between the radial
and the vertical components of the velocity dispersion. Then
\begin{equation}
\label{adrift}
V_c^2-\overline{V}^2=
\sigma_R^2 \left[ \frac{R}{h}-R\frac{\partial}{\partial R}
\ln(\sigma_R^2)-\frac{1}{2}+\frac{R}{2 V_c} \frac{\partial V_c}
{\partial R} \right] , 
\end{equation}
where the disc photometrical scale length $h$ and the radius $R$ are
in arcsec.  By assuming a model distribution for the circular velocity
the observed stellar rotation curve $V(R)$ can be fitted directly using
this equation.  The circular velocity, $V_c$, is modelled as a power
law, $V_c = V_0 R^\alpha$.  If the galaxy contains a cold gas disc,
$V_c$ can be either included in the fitting procedure or serve as a
consistency check on the results. 

The other two
observables, $\sigma_{\rm major}$ and $\sigma_{\rm minor}$, are modelled
assuming exponential distributions for both the radial and the vertical
velocity dispersion components. 
\begin{equation}
\sigma_R = \sigma_{R,0} \exp(-R/a)   
\end{equation}
\begin{equation}
\sigma_z = \sigma_{z,0} \exp(-R/a) 
\end{equation}
Note that there is no a priori reason to believe that the scale length
$a$ is the same for both the vertical and the radial component. Given
the present data, it is not possible to constrain both scale lengths
independently.

There are a total of five free parameters to be determined in our
model.  Three parameters determine the velocity ellipsoid:
$\sigma_{R,0}$, $\sigma_{z,0}$, $a$; and two describe the potential:
$V_0$ and $\alpha$.

\section{Observations \& Reduction}

\begin{center}
\begin{figure}
\epsfxsize=\hsize
\epsfbox{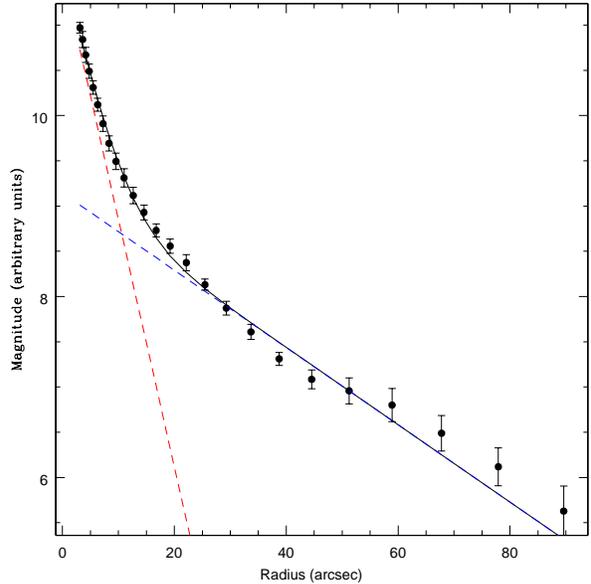}
\caption{Azimuthally averaged radial surface brightness profile of NGC~2985 
($I$-band). Both the disc and the bulge component have been
fitted with an exponential profile. The solid line is the sum of   
the two components.
}
\end{figure}
\end{center}

We have applied the analysis described above to the Sab galaxy NGC~2985. 
It is a typical multi-armed spiral with an inclination of 36$^\circ$
(Grosbol, 1985).  Its distance, using an Hubble constant of 75, is
18~Mpc and the amplitude of the inclination corrected HI rotation curve
is about 250~km/s (WHISP database), see table~\ref{galpar}.

\begin{table}
\caption{Parameters of NGC~2985} 
\begin{center}
\begin{tabular}{lc}
\hline
Hubble Type 	& Sab \\
Inclination 	& 36$^\circ$ \\    
Distance 	& 18 Mpc \\
Max. rotational velocity & $\sim250$ km/s\\
$M_B$ 		& -21.10 \\ 
Disc scale length & 30" in $I$ \\ 
\hline
\end{tabular}
\end{center}
\label{galpar}
\end{table}

\subsection{Spectroscopy} 

Longslit absorption line spectra along the major axis (3.5 hours
integration) and the minor axis (4 hours) of NGC~2985 were obtained
with the ISIS spectrograph on the William Herschel Telescope in the
first week of January 1997.  These spectra were centered around the 
Mg~b feature near 5200 \AA.  Calibrating arc lamp exposures were taken
every 30 minutes.  The dispersion per pixel is $\sim 0.4$ \AA.

The longslit spectra were reduced to log-wavelength bins in the standard
way, using {\sc IRAF} packages.  Adjacent spectra were averaged to
obtain a signal-to-noise ratio of at least 25 per bin, and the
absorption-line profiles of these co-added spectra were analysed. 
Velocity dispersions and radial velocities have been extracted from the
absorption-line profiles using the traditional Gauss-fitting method
after we had first established that the profiles are indeed close to
having Gaussian shape.  The observed spectra are compared with a
template spectrum (HD107288, type K0III) convolved with a Gaussian.  The
Gaussian that best fits the data in a least square sense then yields the
stellar radial velocities and the stellar velocity dispersions. 

The derived velocity dispersions along the major axis are significantly
better behaved (larger extent and less erratic behaviour) on the
redshifted side of the centre than on the blueshifted side.  This
behaviour was also observed with the software of van der Marel (1994). 
It is even apparent in the kinematic data of Heraudeau et al (1999).  We
have therefore decided to only use the redshifted side of the major axis
in the subsequent analysis.  Hence we have only half as many data points
on the major axis as we have on the minor axis, which looks regular on
both sides of the centre.  Around 20 arcsec from the nucleus there is
some indication of a break in the profiles.  Beyond this radius the
velocity dispersion profiles flatten, suggesting that the disc component
starts dominating the velocity dispersions. 

The spectral range of the longslit absorption spectra also included the
5007 \AA \ emission line.  Fitting the mean of the emission line in each
co-added spectrum gives us a direct measure of the circular velocity. 

\subsection{Photometry}

The extent of the bulge can be assessed from a bulge-disc decomposition
performed on a WHT $I$ band image obtained from the ING archive. 
Ellipses were fitted to the isophotes and an azimuthally averaged
luminosity profile was extracted.  The bulge component and the disc
component of this profile were then simultaneously fitted assuming an
exponential distribution for the disc component and either an
exponential profile, Fig.~1, (Andredakis \& Sanders, 1994) or a De
Vaucouleurs profile for the bulge. 

Using a De Vaucouleurs profile gives a slightly better $\chi^2$ value
than using an exponential profile.  The latter, however, is
mathematically more sound since its derivative does not go to zero at
the origin and it is physically more attractive because a De Vaucouleurs
profile will at some (large) radius dominate the total light
distribution again.  With a double exponential distribution the bulge is
significantly smaller than with an exponential disc and a De Vaucouleurs
bulge.  However, even in the latter case the disc dominates beyond a
radius of 20 arcsec (but then only by a factor of a few). 

Taking the average of the two different fitting procedures to represent
the true disc scale length we derive an exponential disc scale length in
the $I$ band of 30 $\pm$ 4 arcsec, close to the literature value of 35
arcsec (Grosbol, 1985). 

\section{Analysis}

\begin{table}
\caption{Best-fit parameters and their corresponding one-sigma errors
obtained from a brute force calculation.
The errors in brackets are obtained from bootstrapping the data.}
\begin{center}
\begin{tabular}{lc}
\hline
Parameter & Best-fit  \\
& \\
$\sigma_{R,0} (\kms)$ & 149	  $\pm$ 12   (15)   \\
$a$ (arcsec)          & 73	  $\pm$  9   (15)   \\
$\sigma_{z,0} (\kms)$ & 127	  $\pm$ 10   (15)   \\
$V_0 (\kms)$          & 136	  $\pm$ 14   (28)   \\
$\alpha$	      & 0.18	  $\pm$ 0.03 (0.07) \\
\hline
\end{tabular}
\end{center}
\label{fitpar}
\end{table}

\begin{center}
\begin{figure*}
\epsfxsize=\hsize
\epsfbox{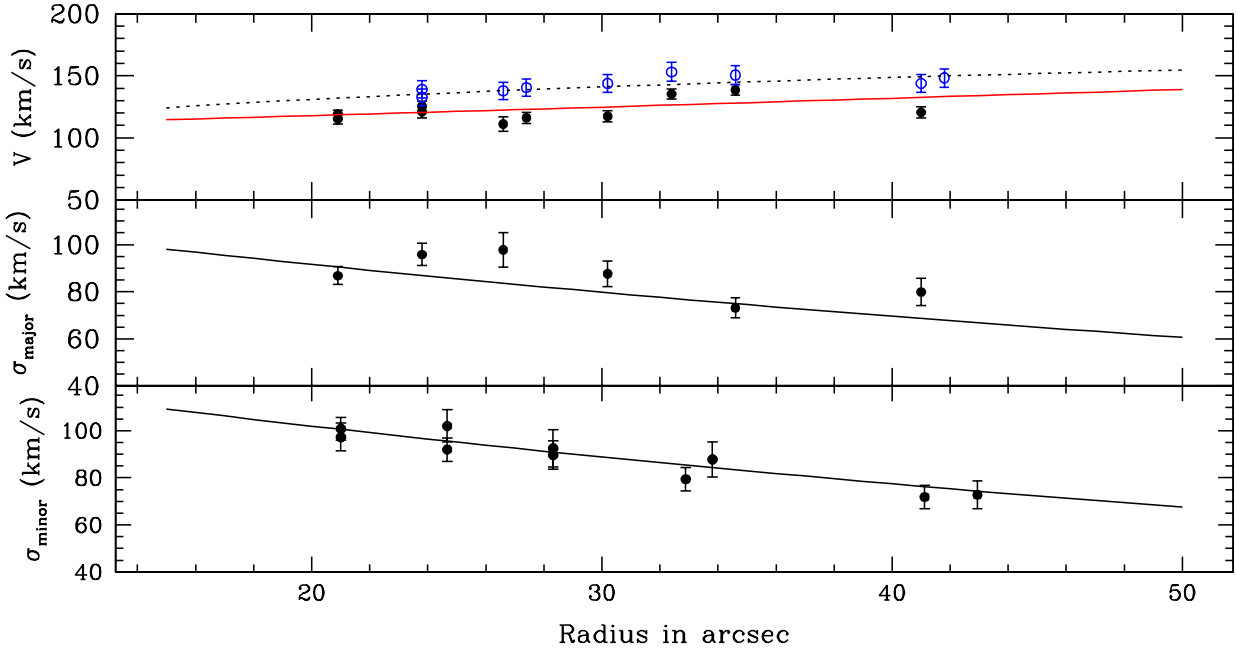}
\caption{
The best-fitting model distributions (solid lines for the stellar
kinematics and a dotted line for the emission line kinematics) and the
line-of-sight velocity data.  Within 20 arcsec the bulge dominates the
total light distribution.  This region has therefore been excluded in
the analysis.  The open circles in the upper panel are measurements of
the circular velocity determined from the emission line at 5007 \AA \
while the filled circles measure the stellar rotation derived from the
absorption lines. 
}
\label{bestfit}
\end{figure*}
\end{center}

The projected model distributions of section 2 were fitted
simultaneously to the observables.  However, unlike in Paper 1 we were
unable to apply the non-linear fitting (Press et al.  1992) routines
successfully, probably because the present data set is noisier and
contains fewer data points.  Fitting is therefore accomplished using a
combination of a simulated annealing (a.k.a.  biased random walk, Rix \&
White 1992) method and the downhill simplex method (Press et al.  1992). 
Since the number of data points is rather small we have included the
emission line data directly into the fitting procedure.  This allowed us
to better constrain the model distributions although it also meant that
we had to sacrifice the consistency check on the result. 

\begin{figure} 
\epsfxsize=\hsize 
\epsfbox{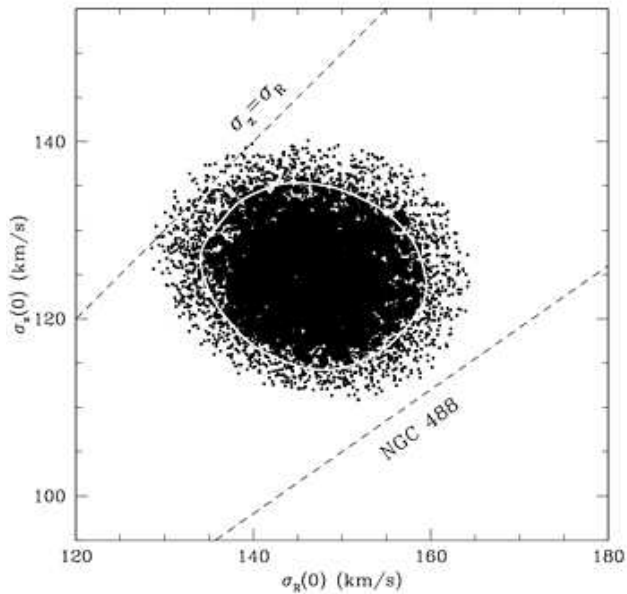} 
\caption{ 
Distribution of the allowed range of values of the extrapolated
central radial and vertical velicity dispersions. Small dots are
random choices of the model paramaters (projected onto $\sigma_R(0)$,
$\sigma_z(0)$) which are consistent with the
data at the 2-$\sigma$ level, large dots are consistent at 
1-$\sigma$. The white contour delineates the 1-$\sigma$ region.
The upper dashed
line indicates an isotropic $\sigma_z=\sigma_R$ velocity ellipsoid.  The
results obtained for NGC488 (Paper 1) are also indicated. 
}
\label{sigrsigz}
\end{figure}

Error estimates were obtained numerically since the minimisation of a
function does not provide direct estimates of the uncertainties
involved.  One-sigma errors have been estimated from a brute force
calculation.  In this procedure we calculated $\chi^2$ values by varying
all five fitting parameters in a broad range around their best-fit
values ({\it i.e.} $\chi^2$ is sampled on a five-dimensional grid). 
One-sigma errors correspond to an increase of unity in $\chi^2$ over its
best-fit value.  The projection of the grid inclosure where $\Delta
\chi^2 = 1$ onto the principal axes determines the one-sigma uncertainty
in each individual parameter.  A two dimensional projection of this grid
is shown in Fig.~\ref{sigrsigz}.  Here the grid is projected onto the
plane spanned by the radial and vertical velocity dispersion axes.  The
distribution of points, `the error cloud', not only yields the one-sigma
errors but it also provides a measure of the covariance between the two
parameters. 

The confidence intervals have also been estimated from a Monte-Carlo
study of a large number of new data sets, with each new data set drawn
randomly from the original data set.  This bootstrapping procedure
resulted in slightly larger error-bars but did not change the numerical
values of the parameters suggesting that the brute force calculation is
probably a bit to restrictive in its estimate of the uncertainties.  For
the two parameters that we are most interested in, $\sigma_R$ and
$\sigma_z$, the difference appears to be at a minimum anyway. 

The obtained best-fit parameters are listed in table~\ref{fitpar} and
the corresponding model distributions are together with the data
presented in Fig.\ref{bestfit}.  The associated $\chi^2$ value of 50,
however, is a bit higher than what is formally required by chi-square
fitting.  But the scatter of the data points around the best-fit models,
especially along the major axis velocity dispersion is rather large and
the large value of $\chi^2$ can probably be attributed to that.  It even
appears that the best-fit profile along the major axis actually lies too
low given the data points.  Along this axis the observed dispersions are
a combination of the tangential and the vertical velocity dispersion
components while the minor axis measures the radial and vertical
components.  Implying that the dispersions along the minor axis should
be higher than along the major axis if the components are distributed in
the canonical way, $\sigma_R > \sigma_\phi > \sigma_z$. 
 
However, the difference between the circular velocity and the stellar
rotation speed (top panel of Fig.  2) is roughly proportional to the
radial velocity dispersion, see equation~\ref{adrift}.  A higher
best-fit major axis profile therefore also implies a smaller difference
between the circular velocity and the rotational velocity, which is
certainly not warranted by this data set.

The best-fit kinematical scale length parameter, $a$, is about twice as
large as the photometrical scale length.  A result predicted by the local
isothermal approximation of stellar discs ({\it e.g.} van der Kruit \&
Freeman, 1986).

\section{Discussion}

The results derived in the previous section imply a ratio of the
velocity ellipsoid of $\sigma_z/\sigma_R$ = 0.85 $\pm$ 0.1.  The
one-sigma error includes the covariance -- evident from the fact that
the distribution of points in Fig.~\ref{sigrsigz} is not completely
circular -- between the two parameters.  With the error estimates
obtained from bootstrapping the ratio becomes 0.85 $\pm$ 0.13. 

All the one-sigma points in Fig.~\ref{sigrsigz} are located below the
line $\sigma_z = \sigma_R$.  Indeed, according to the generic picture of
the gradual heating of stellar discs (Jenkins \& Binney, 1990) the ratio
$\sigma_z/\sigma_R$ -- starting from an isotropic distribution of
velocity dispersions -- will become smaller than one given enough time. 
If only giant molecular clouds are responsible for heating this ratio
will approach 0.75 and, if spiral structure also contributes to the disc
heating the ratio will be lower.  The derived ratio is therefore just
consistent with the picture where disc heating is dominated by giant
molecular clouds.

\subsection{Comparison to other galaxies}

\begin{figure}
\epsfxsize=\hsize
\epsfbox{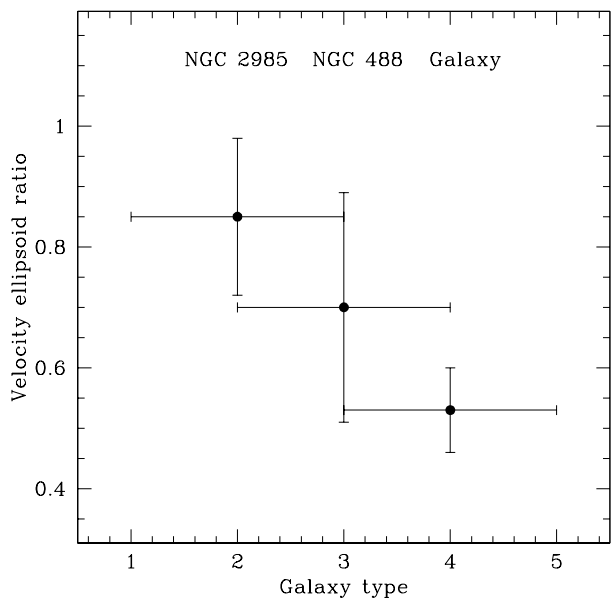}
\caption{
Observed velocity ellipsoid ratios as a function of morphological RC3 type. 
The observed ratios in both NGC 2985 and NGC 488 are average values at
one disk scale length while the value derived for the Galaxy is obtained
at two disk scale lengths. The horizontal error bars reflect the uncertainty
associated with classifying galaxies. 
}
\end{figure}

The obtained velocity ellipsoid axis ratio is a little higher but
comparable to the value found for NGC~488 ($0.70 \pm 0.19$).  The
slightly rounder velocity ellipsoid found in NGC~2985 may reflect a
genuine difference in the internal heating mechanisms.  Since both galaxies
are multi-armed type spiral galaxies but the Sab galaxy NGC~2985 is of
slightly earlier type than NGC~488, which is an Sb galaxy. 

However, in both galaxies the axis ratios (average values around one
disc scale length) are somewhat higher than in the solar neighbourhood
(at about two disc scale lengths) where the most accurate measurements
using the {\it Hipparcos} data indicate a ratio of 0.53 $\pm$ 0.07
(Dehnen \& Binney 1998). 

This behaviour is again consistent with the predictions
of simple disc heating mechanisms since our own Galaxy is of type Sbc
and therefore has a higher contribution from heating by spiral structure
resulting in a flatter velocity ellipsoid. 

This effect is illustrated in Fig.~4.  The observed trend may be
slightly fortuitous given the large error-bars on each point.  However,
photometric observations of edge-on galaxies (van der Kruit and de
Grijs, 1999 and references therein) indicate a similar trend, {\it i.e.}
late type spiral galaxies are more flattened than early type spiral
galaxies. 
 
A larger sample is clearly needed to clarify the significance of this
trend.  The relevant measurements for a single galaxy typically require
one night on a 4m class telescope.  Doubling or tripling the sample size
can thus be done in a relative short time span and should unequivocally
establish the validity of the current disc heating theories. 

\section*{ACKNOWLEDGEMENTS}

The WHT is operated on the island of La Palma by the Isaac Newton Group
in the Spanish Observatorio del Roque de los Muchachos of the Instituto
de Astrof\'isica de Canarias. Much of the analysis in this paper was
performed using {\sc iraf}, which is distributed by NOAO.

\end{document}